# The Future Internet of Things and Security of its Control Systems

Misty Blowers, USAF Research Laboratory
Jose Iribarne, Westrock
Edward Colbert, ICF International, Inc.
Alexander Kott, US Army Research Laboratory

## Introduction

We consider the future of ICS cyber security. As best as we can see, much of this future unfolds in the context of the Internet of Things. In fact, we envision that all industrial and infrastructure environments, and cyber-physical systems in general, will take the form reminiscent of what today is referred to as the Internet of Things.

Internet of Things is envisioned as multitude of heterogeneous devices densely interconnected and communicating with the objective of accomplishing a diverse range of objectives, often collaboratively. One can argue that in the relatively near future, the IoT construct will subsume industrial plants, infrastructures, housing and other systems that today are controlled by ICS and SCADA systems.

The advent of IoT will be accompanied by a number of developments: miniaturization of devices and sensors, increasing mobility of devices, wearable devices, ubiquitous robotics and growing automation of all functions of IoT. Many of these devices will be smart sensor that contains a microprocessor that conditions the signals before transmission to the control network. Some of the devices are likely to be nano-robots with overall size of the order of a few micrometers or less in all spatial directions and constituted by nanoscopic components.

IoT will be associated with great increase in automation. In addition to supporting highly autonomous devices, IoT itself will be self-organizing, self-configuring, and self-healing. The increase in automation may cause an increase in system vulnerability. With automation comes the necessity of reducing the need for manual intervention. Automated security monitoring will be essential as control systems grow large enough to exceed the capacity for humans to identify and process security logs and other security information.

Other game-changing development may include radically new computing and networking paradigms. Emerging computing paradigms—nanocomputing, quantum computing, biologically or genome-based computing—might develop soon enough to make most current cybersecurity technologies obsolete, thus drastically changing the market. Quantum computing and networking are already fueling lively debate. Biologically inspired computation and communication paradigms will attract growing interest, especially as they offer promises for autonomous adaptation to previously unknown threats and even self-healing.

In the IoT environments, cybersecurity will derive largely from system agility, moving-target defenses, cyber-maneuvering, and other autonomous or semi-autonomous behaviors. Cyber security of IoT may also benefit from new design methods for mixed-trusted systems; and from big data analytics—predictive and autonomous.

## Overview of Change in Control Systems

### Industrial Revolution: Earliest Times to the Present

The first industrial revolution began in Britain in the late 18th century, with the mechanization of the textile industry. Tasks previously done by hand in hundreds of weavers' cottages were brought together in a single cotton mill, giving birth to the factory. The second industrial revolution came in the early 20th century, when Henry Ford improved the moving assembly line and ushered in the age of mass production (The Economist 2012). There is a debate regarding electrification and electronics, including automation, being a possible third industrial revolution leading into a fourth. It is clear that a major change is now underway; manufacturing is becoming digital. The modern world is seeing the convergence of the global industrial systems with large-volume data capture and analysis, all enabled by ever increasing computing power. The distributed growth of networked systems, internet connectivity, low-cost wireless technology, advanced sensors, and satellite systems are shaping a new world where the reliance of man on machine is dominant.

Industry or manufacturing (we will use these terms interchangeably) is largely the process of conversion of raw materials into products. Manufacturing is increasingly dependent on sophisticated equipment and automation to meet simultaneous demands for safety, quality, efficiency and productivity. However, different generations of equipment and automation co-exist as older plants and mills, or different production areas therein, and continue to operate along their more efficient and newer brethren. Increasingly, the distinction between equipment and automation is becoming blurred as new process equipment has embedded sensing, control and communication devices.

According to the US President's Council of Advisors on Science and Technology, advanced manufacturing is "a family of activities that (a) depend on the use and coordination of information, automation, computation, software, sensing, and networking, and/or (b) make use of cutting edge materials and emerging capabilities enabled by the physical and biological sciences, for example nanotechnology, chemistry, and biology. It involves both new ways to manufacture existing products, and the manufacture of new products emerging from new advanced technologies." (Holdren et al. 2012)   Additional studies have shown, however, that there is a growing gap between research and development activities and the deployment of technological innovations. There is a recognized need to accelerate the technology life-cycles in the U.S., and growing numbers of entrepreneurial programs are enabling this to happen.  The acceleration of the technology life-cycle increases the importance of gaining market share in the commercialization phase so that manufacturers can seize the opportunities associated with the scale-up phase.

These changes will come with a cost, however.  The faster we push these technologies into the manufacturing environment, the higher the risk and potential for failure.  Economic gain will be realized with evolutions of core products, but the biggest gains will come from the disruptive technologies that can revolutionize current methods or products.

## Sustainability of an Industrial Enterprise

In the manufacturing context, sustainability is essential to the long-term survival of an enterprise constrained by economic, environmental and social factors. Those are primary considerations for investments in new technology.

### *Economic Factors*

The economic constraints of a modern company include the escalation and volatility of material and energy costs, customer and market pressures to accelerate new product introductions and the continual push for greater productivity and cost reduction. Companies also face the escalation of capital expenditures, as modern equipment is increasingly more costly to purchase and install. Companies also face the inevitable obsolescence of equipment that is still productive but contains parts that are no longer manufactured,. This is especially true of ICS, where the trend towards the use of "commercial off-the shelf" computer hardware has reduced initial costs, but also shortened the life expectancy of the computers as their operating systems become unsupported every few years.

*Environmental Factors*

The environmental pressures on manufacturing include increasingly tighter regulations for emissions to the air and water, and waste generation, as well as concerns over global climate change. In response, many companies have adopted targets to reduce their carbon footprint, *i.e.* the direct and indirect emissions of carbon dioxide associated with their operations. The environmental constraints are most acute in industrial operations dealing with dangerous substances and hazardous processes, especially in the chemical and nuclear industry. Major accidents with multiple fatalities continue to occur worldwide in the process industries, causing distress to those affected and massive costs to companies. Accidents at Flixborough, U.K., Seveso, Italy, Bhopal India, and Pasadena, Texas, in the 1970s and 1980s led to tighter regulation of the process industries and raised awareness of the key risk control systems needed to prevent such accidents (Kletz 2009). In the United States, companies need to comply with both the OSHA Process Safety Management and EPA's Risk Management Program. Those rules require a process hazard analysis to be conducted and risks to be reduced to an "as low as reasonable practical" level. Similar regulations exist in other countries and in most cases require inherently safe (Moore 2006) or instrumented safety systems, including a hierarchy of controls and redundancy. The current rules for such systems generally do not allow for internet and wireless technology, seriously limiting the adoption of IoT technology.

*Social Factors*

A major social constraint on manufacturing, at least in developed countries, is the aging of the technical workforce. Employers find that replacing qualified workers and engineers is increasingly difficult as they retire. In the U.S., the median age of the manufacturing workforce spiked to 46.1 years in 2013, up from 40.5 years in 2000. For high-skilled manufacturing workers, the average age is 57 (Higgins 2015). Cavallaro (2015) cites a study by Deloitte and The Manufacturing Institute that illustrates just how dire the situation has become: six out of ten manufacturing positions remain unfulfilled because of the talent shortage, and the projected shortfall may rise to two million workers in the next decade. And yet, 52% of American teenagers have no interest in a manufacturing career.

*The Future*

The most likely method for industrial sustainability will be increasing the degree of automation of the manufacturing processes. For example it is possible to reduce the required personnel in assembly lines by up to 90% through the use of robotics (Forrest 2015). The remaining workforce will need to be highly skilled and better trained to compensate for the smaller number of employees (Young 2015). Outsourcing is another possible solution. Outsourcing allows in-house personnel to focus on day to day priorities, while the less critical work is performed by contractors. Due to the spread of the Industrial Internet of Things with its non-proprietary character, the major suppliers of automation are taking a defensive position by offering service agreements that typically include condition-based monitoring, remote troubleshooting, spare parts and technical labor.

As Industrial enterprises include more and more automation, for example in the forms of robotic hardware and smart machines, the number of vulnerable paths through which the adversary may exploit system processes increases dramatically. This happens almost unnoticeably since industrial operators and control system builders are not typically focused on security aspects during design, construction and testing. The types of vulnerabilities can become extremely diverse as a plant converts process elements to uniquely manufactured automated devices customized for that particular process element. Software and firmware vulnerabilities grow to offer a much greater attack surface than is currently available to the dedicated adversary. Plant operators and owners will need to increase their security staff or hire specialized security analysts to accommodate the deteriorating security of their systems. Alternatively, vendors could offer more secure hardware, firmware, and interconnections. This is less likely to occur in the short term.

## The Internet of Things (IoT)

The term "Internet of Things" (IoT) and "Industrial Internet of Things" (IIoT) describe a vast number of connected industrial systems that are communicating and coordinating their data analytics and actions. As ICSs evolve, IIoT devices and methods will be introduced to improve industrial performance. Industrial systems that interface the digital world to the physical world through sensors and actuators that solve complex control problems fall under a much broader category of "Cyber-Physical Systems" (Monostoria 2014). The term "Cloud Manufacturing" (Wu et al. 2013) describes the distributed or remote infrastructure that will likely be needed to handle the growing amounts of information and demands on computer processing speeds in the manufacturing facilities of the future.

Although these advances were forecasted by several authors in the early 1990s, notably by Mark Weiser (Weiser 1991), interest in the integration of advanced digital technologies into industrial production systems did not spread until the following decade, when related industrial consortia and governmental initiatives were started in several countries.

### Global Development of the IIoT

A non-profit registered association named "Technology Initiative SmartFactory" was established in Germany in 2005 to develop, apply and distribute innovative, industrial plant technologies, and to create the foundation for their widespread use in research and practice (Zuehlke 2010). The partner circle grew rapidly, including producers and users of factory equipment as well as universities and research centers. Support was provided by industry and political organizations and eventually became national German policy as part of the "Industrie 4.0" plan, first discussed in 2011 and later adopted in 2013. The heart of the Industry 4.0 idea is intelligent manufacturing, *i.e.* applying the tools of information technology to production. In the German context, this primarily means using the IIoT to connect small and medium-sized companies more efficiently in global production and innovation networks so that they could more efficiently engage in mass production and more easily and efficiently customize products (Kruger et al. 2014).

The IIoT development efforts in Europe are being monitored by The Internet of Things European Research Cluster, which maintains its Strategic Research and Innovation Agenda (SRIA) taking into account its experiences and the results from the on-going exchange among European and international experts. The SRIA is updated every year with expert input from projects financed by the European Commission (Vermesan & Friess 2013).

In the United States, several private-industry consortia were formed starting with the "Object Management Group" in 1989 and have taken a leading role in developing standards for the IIoT. Relevant consortia include the "Data Distribution Service," the "Smart Grid Interoperability Panel" and "Open Interconnect." Particularly important is IIC, the "Industrial Internet Consortium," started in March 2014 by AT&T, Cisco, General Electric, IBM and Intel. IIC now has more than 200 member companies from 25 countries and recently released its reference architecture for the industrial internet (Industrial Internet Consortium 2015). Through the National Institute of Standards and Technology, the Federal government started an Advanced Manufacturing program that includes many technologies related to IIoT. One of the program objectives is to create several linked institutes for manufacturing innovation, with common goals but unique concentrations (NIST 2015). For example a new 94,000 square feet Digital Manufacturing and Design Innovation Institute / UI Labs opened on May 22, 2015 in Chicago, Illinois.

"Made in China 2025" is a plan released in May 2015 to comprehensively upgrade Chinese industry (Kennedy 2015). The initiative draws direct inspiration from Germany's Industry 4.0 plan, but the Chinese effort is far broader. Its guiding principles are for manufacturing to be innovation-driven, to emphasize quality over quantity, achieve green development, optimize the structure of Chinese industry, and nurture human talent. The goal is to comprehensively upgrade Chinese industry, making it more efficient and integrated so that it can occupy the highest levels of global production chains.

### Expected Impact

It is widely expected that the IIoT will have an enormous impact. Its global economic added value has been variously estimated between $1-trillion and $20-trillion of GDP growth in fifteen years (Press 2014). However, the introduction of the IIoT is expected to occur more gradually and be less disruptive than previous industrial revolutions. According to a report by McKinsey & Co (Meister 2015), the implementation of the IIoT will require the replacement of 40 to 50 percent of the current equipment in traditional industries. Those figures compare favorably with the introduction of industrial automation, which required an 80 to 90 percent rate of replacement.

As mentioned in the previous section, the increase in automation will cause a major increase in system vulnerability until security measures are included with the new hardware and software, and security staff is increased appropriately to monitor the new "things," services, and methods in the IIoT. Automated security monitoring will be essential as control systems grow to exceed the capacity of humans to identify and process security logs and other security information.

## Game Changers in the Future ICS and IoT Security

In this section we explore specific aspects of present and future control systems that we believe will greatly affect the design and security of future ICSs and the IoT as a whole As shown in **Error! Reference source not found.**, we group the relevant aspects into three general areas:

1. Construction of the Future IoT – commercial and academic efforts to design and build innovative new "things" that other will use

2. Users of the Future IoT – industrial users and consumers who use these "things"

3. Support for the Fugure IoT  -  services and collaborative efforts to support the ability of users to use the new "things"

We identify aspects in each of these general areas that will contribute significantly to the future security of ICSs.  As mentioned, industrial control is not limited tomanufacturing and other industrial processes.  Devices and controllers are used by consumers worldwide for controlling home lighting, security cameras, automobiles, and many more home-based sensors.  One difference is that home-based devices and controllers are cheaper, are mass-manufactured with generally poor software and firmware security, and are usually connected to the Internet.  ICSs were designed with the general understanding that they would have no network connectivity to the outside world.  However, this is changing as industry wishes to exploit the advantages in convenience provided by expanding network connectivity.

As shown in **Error! Reference source not found.**, we break our three general areas into eight categories.   In the following sections, we elaborate on specific aspects of these eight categories.

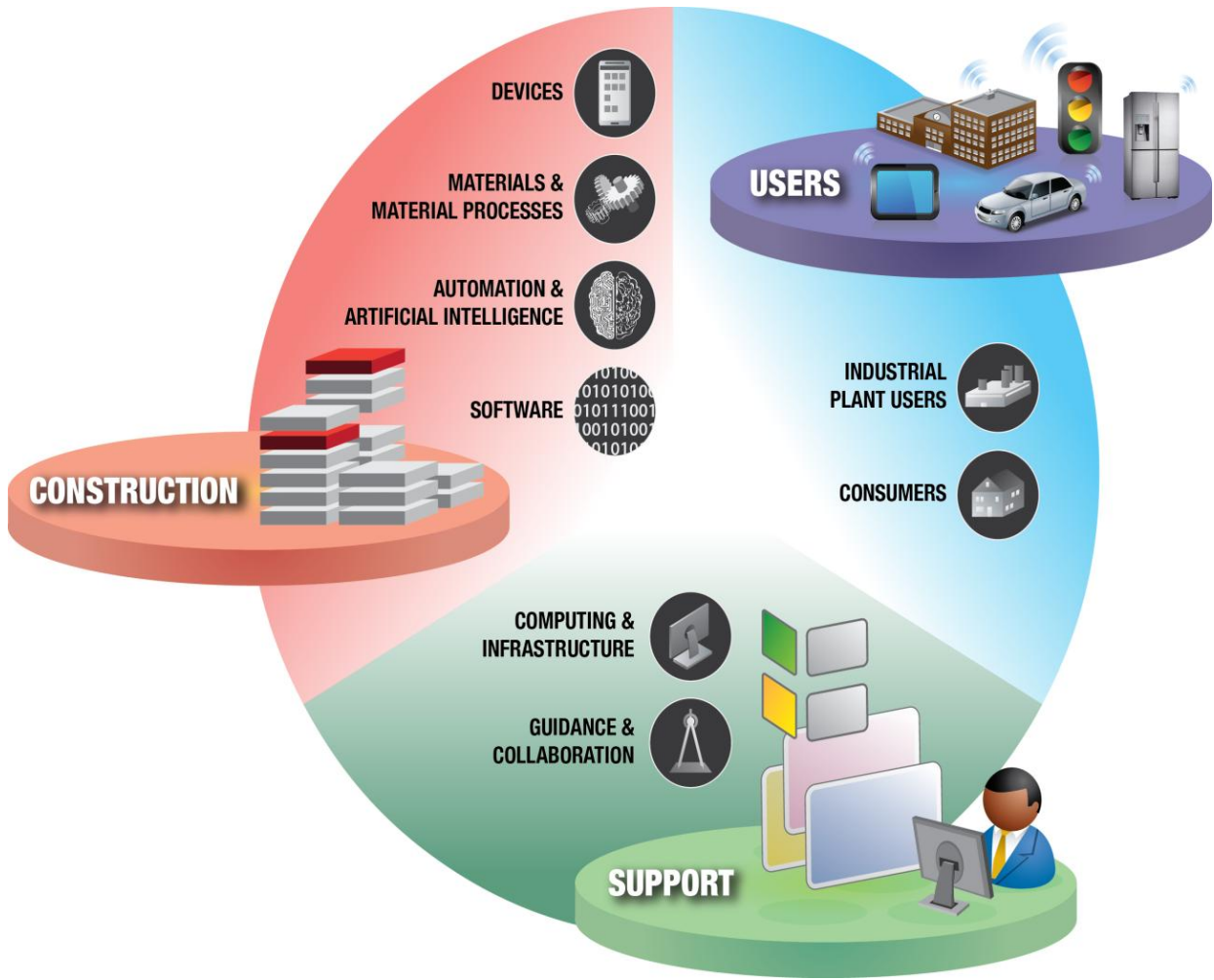

Eight Categories within IoT Focus Areas:

- Focus Area 1: Construction of the Future IoT
  - Devices
  - Materials and Material Processes
  - Automation and Artificial Intelligence
  - Software
- Focus Area 2: Users of the Future IoT
  - Industrial Plant Users
  - Consumers
- Focus Area 3: Support for the Future IoT
  - Computing and Infrastructure
  - Government and Industry Guidance and Collaboration

## Construction of the Future IoT
*Devices*

### Miniaturization of End Devices and Sensors

As transistor density on silicon-based chips continues to follow Moore's law and doubles every 1.5-2 years, not only does overall computing ability increase, but computing ability per unit volume increases. Hand-held devices of today have the computing power of "supercomputers" of yesteryear. For the future IoT, this has a number of important implications.

Miniaturized computing devices will be more ubiquitous due to mass manufacturing at relatively low cost. This includes not only CPU and memory chips, but RF and other sensor-based technologies integrated into System-on-a-Chip technologies. Innovative packaging methods for chips (e.g., Charles 2005) will allow considerable flexibility for future manufacturers and integrators.

With more computing power in miniature computing devices comes a requirement to process and condense larger amounts of sensor and other data being processed by the end devices. Proxy communication by cellular phones is currently being used between miniature end devices (such as wearable fitness devices) and the Internet and cloud storage, as there is no possibility to store all of the sensor data in the cloud (e.g. Want, Schilit, and Jenson 2015). Endpoint devices must be able to pre-process raw data and forward a useful subset of the information to the proxy or directly to the Internet. Proxy devices must be able to handle the volume of the network traffic and communicate safely and reliably to the end devices. Browser protocols such as HTML for human-to-machine (H2M) interaction will need to be updated with machine-to-machine (M2M) protocols for increased efficiency (Want et al. 2015).

In order to accommodate local network traffic, cloud storage models may favor increased amounts of storage and processing in local servers such as cloudlet servers, which could run as virtual machines on desktops or even dedicated embedded servers.

Securing the array of miniaturized devices will be challenging initially (cf. Green 2015). This is mostly due to the fact that the driving force in the IoT is marketing of the new technologies, not the security of the new devices. Inexpensive devices that can increase profits of a company are highly attractive to corporate decision makers, even if a security risk is implied.

The number of embedded devices and sensors will increase, by a factor of ten, and even higher. Some will have IP addresses and will communicate with secure TCP protocols via secure applications, but many will use proprietary or ad-hoc communication methods, such as insecure 802.11 or Bluetooth wireless. The data from the sensors and devices will be accessible from the proxy server, which may be a cellphone or a small dedicated embedded device in an industrial setting. Access to that proxy server can provide an adversary with the ability to inspect or modify a much larger amount of information than before. To preserve confidentiality and availability, system owners should analyze the information being recorded and communicated by the end devices and sensors, and protect access as needed. At some point, as with the Internet, the amount of information will be too large for a human to monitor, and security tools will need to be developed to validate secure data flow from end devices and sensors.

As the number of end devices and sensor, and proxy devices increases, so does the volume of software that controls those devices. Software is developed by humans and always has vulnerabilities than can be exploited, especiallyt if innovative end devices are pushed to market quickly with little security engineering. Unauthorized access to information in proxy servers or end devices themselves will need to be analyzed and vulnerabilities will need to be mitigated. While it would make sense for this activity to be preformed before or during installation of IoT devices, it is often neglected until after an incident is reported.

## Mobility and Wearable Devices

Recent advances in battery life, miniaturization, energy harvesting, communication protocols, and lower hardware costs are bringing the vision and utility of the IoT closer to reality (Zorzi et al. 2010). Mobile devices and wireless devices connected to mobile networks are key aspects of this development process. The number of internet connected devices has already exceeded 1.0 per person on the planet, and is expected to be 4.3 per person by 2020 (Waring 2014). The wearable technology market is expected to grow by a remarkable factor of three in the next three years (2015 to 2018, Rizzo 2013). By 2020, there is expected to be an additional 2 devices per person on the planet, including end devices, sensors, and wearable devices. How will the current communication infrastructure handle this additional burden?

Mobile devices will impact the IoT infrastructure in a number of ways. Many personal-based wearable devices do not connect directly to the internet or to a corporate network, but connect to a mobile device such as a cellular phone that serves as a communication hub for the wearables. In addition, the lack of a direct connection to the wearable devices offers some privacy security to potentially sensitive information. While RFID devices such as identification badges, credit cards, and passports may not currently be integrated into the consumer-based IoT, their utility as authentication of identity and location in the workplace can easily fit into the future global IoT. For example, as one maneuvers through a physical plant or one's home, it may be desirable for the lighting, HVAC system, audio-visual systems, or other "things" in the workplace or home, to recognize one's presence and adjust accordingly. For a home setting, one might always want the television to resume a video series with the window shades and lighting adjusted accordingly.

For third-world countries (Glickman 2015), mobile phones provide crucial news and agricultural information so that small-scale farmers can plant and harvest food more effectively based on weather information, seed prices, and market demand. Many countries have poor or non-existent wired infrastructure, so inexpensive mobile connectivity offers a great utility for improving agricultural efficiency.

First-world country industry and health-care are not entirely dissimilar. Manufacturing plants save tremendous costs if remote sensing data from "things" can be placed anywhere in the plant and the data fed back wirelessly. Plant operators with wearable technology will provide crucial feedback about the plant environment as they visit locations within the plant during the day. Wearable devices have the potential to make operators more mobile and effective. Devices like the virtual reality headsets allow operators to have a more global view of plant operations and are invaluable for training exercises. Augmentation of the human body can increase human strength for lifting heavy objects (Hirukawa 2015).

Hospital workers already use IoT methodologies. Sensors in rooms identify humans and material assets for inventory and emergency purposes. Medical sensors for blood pressure, pulse, oxygen level, and other vital statistics can report the information directly to a central database, which is readily accessible by clinicians by laptop or tablet via wireless communication. Privacy concerns are significant, as HIPAA laws strictly protect personal information.

This privacy concern also applies to health-related wearables in the consumer market. The consumer will want to share vital health statistics with health-care providers, friends, and family, but not with general public who may have physical proximity to the wearable. In addition, wearables with cameras or microphones, such as Google Glass, have the potential of violating the privacy of others by recording audio or video. While this is not a new problem, broad use of the Glass has resurfaced the issue.

Consumer-based smart watches and fitness trackers are increasingly becoming fashion accessories. Wearable device use will soar. Aesthetics of smart phones and miniature mobile devices have always been important in the consumer market.

How will the global IoT accommodate the expected exponential growth of mobile device connectivity? Adaptation of heterogeneous access network and efficient use of available resources are important. Large numbers of mobile devices with multiple tethered wearable (or local) devices will be roaming in and out of mobility cells in automobiles, trains, airplanes, and drones. Machine-to-Machine (M2M) communication is an important facilitating technology for the IoT, and future M2M communication methods need to accommodate this expanding demand for connectivity. Methods from Heterogeneous MANET (Mobile Ad Hoc Network) (Ahmand et al. 2015) may be useful in this regard.

Until there is better guidance on privacy of personal information (not just health-care related information), and better security guidelines on wireless communication methods and data/cloud storage, security of wearable and other mobility devices will remain poor. Eventually many of the wearable devices that are tethered to cellular and mobile phones will be released with automated internet connectivity to IPv4 and/or IPv6 networks, which will allow them to be publicly accessible. Within an industrial setting, this would mean that access and authentication vulnerabilities would be available to any adversary that gains physical access to the wireless signal. Since many of the end devices would have automatic authentication to the network (or to the operator mobile phone), adversaries would have a much larger number of attack vectors than before. Most likely, the end devices would not have been hardened. Initially, the mobility-based IoT will be very vulnerable to attack. Careful analyses of the control system networks and devices should be done, and appropriate mitigations should be put in place.

*Materials and Material Processes*

### Advances in Materials

Materials are the building blocks of every physical product. Improvements to materials such as steels, metals, plastics, and ceramics have been vital to many of significant technological developments. The newer nanoscale, biological, smart, and composite materials will enable future technological breakthroughs. Some of these breakthroughs will transform existing industries while others will spawn entirely new ones. (Holdren et al. 2012)  The advances in material science are co-evolving with advances in 3D-Printing. The demand for new material properties is partly driven by what is feasible with a 3-D Printer. However, imagine a scenario where a malicious actor "hacks" into your 3-D printer and steals critical design plans? What if the hacker tampers with the design just enough to impose a flaw to the structural integrity to a printed component for an aircraft? We explore security concerns with 3D-printers below.

Advanced materials offer the potential to make vehicles much lighter, dramatically increase the energy density of batteries, or allow a much lighter alternative to glass in space based systems.  Consider concrete as one example. It is difficult to imagine just how much concrete exists in our manufacturing facilities and roadways worldwide, but it is undeniable that Concrete is a very prevalent material in manufacturing facilities and roadways worldwide. Its use, however, is limited by its inherent susceptibility to cracks, and leaks due to the fact that concrete often develops micro-cracks during the construction process. Although these tiny cracks may not immediately affect the building's structural integrity, they eventually can lead to leakage problems. Leakage can eventually corrode the concrete's steel reinforcements, which can ultimately cause a collapse. With the emerging self-healing technology, cracks can be sealed immediately, preventing future leakage and the high cost of repair. (Matchar 2015).

Self-healing materials are inspired by the healing mechanisms of the human body.  Self-healing concrete works by embedding capsules of limestone-producing bacteria and calcium lactate within concrete. When the concrete cracks, air and moisture cause the bacteria to begin consuming the calcium lactate. They convert the calcium lactate to calcite, an ingredient in limestone, thus sealing off the cracks (Matchar 2015). The bacteria can lie dormant for as long as 200 years, well beyond the lifespan of most modern buildings.

So how could something like this be a cyber security concern?  Here the supply chain vulnerability is a major component of security. It is becoming an increasingly greater concern as the logistic chains for even some of less noteworthy components of manufacturing processes often cross international boundaries.

Imagine a scenario where a hacker interferes with the supply chain of this "self-healing" concrete.  What if the supply chain is contaminated in a manner to allow the bacteria to continue after   consumption of the calcium lactate? How much security do we need to consider, not only in our own manufacturing facilities, but also in the facilities which supply raw materials to us?

3D Manufacturing

3D Manufacturing is very much connected to advances in material science. Advances in printing technologies have opened the potential for conformable electronics and physical components and even for subsystems and components embedded in 3D structures. Over the past 20 years, 3D additive manufacturing technologies have been advancing at a rapid pace. These systems have been used in a variety of applications ranging from conventional prototyping and rapid tooling to more advanced applications such as medical implants, aerospace and automotive manufacturing, 3D electronic devices, and micro-systems (Pique et al. 2006, Melchels et al. 2012). The technologies are becoming more accurate with features ranging from micron-sized to building sized (Joshi et al. 2012). The process removes the traditional limits on part geometry, and leads to components that can be produced faster while consuming less material and using less energy.

Precision modeling and simulation may be combined with additive manufacturing to create complex parts that are impossible to manufacture today. Features like durable lattice work, intricate textures and organic shapes are all possible, and even extensions and optimization of existing component parts have been made possible with 3D printing technology. The reduction of mass of printed devices can lead to vast improvements. For example, 3D printing reduced the mass of an antenna-reflector from 395g to around 80g (Williamson 2015).

In spite of its benefits, 3D printing raises concerns from the security perspective. Indeed, engineers of the future will need to have knowledge of cyber security. Advances in software tools that provide automatic correction for 3-D printing does offer some potential to protect 3D printing processes from hacks and from model corruption. However, with every "auto-correction" software tool, there is the potential for an "auto-corruption" tool. Also, as previously mentioned, there is potential for a malicious actor to reside on your system or network, learning about what you printing, or what blueprints a supplier may be transmitting to the end user or customer. The 3D printing technologies are susceptible to all the "D5 effects" (deception, denial of service, disruption, degradation, and destruction).

*Automation and Robotics*

Automation and Artificial Intelligence

As the number of end devices and sensors increases in the IoT, these will be utilized to reduce operating costs or to increase process efficiency. A reduction of manual processes and an increase in automated processes will be a significant benefit to the consumer and to industrial IoT. A goal is for the industrial IoT be "self-organizing, self-configuring, self-healing, scalable to large sizes, with very low energy consumption, low cost, simple to install and based on global standards." (Pinto 2012). In this vision, vendors will work together so that addition of new sensors or new software or networks will be handled automatically, with no manual effort required. The current lack of hardware and communication interoperability presents a significant challenge to overcome before such advanced automation can be realized.

Increased automated feedback from the increased number of sensors and higher fidelity of those sensor readings can provide great value in an industrial setting. Automatic analysis of the data and dynamic adjustments in the process can lead to major reductions in waste, energy costs, and human intervention (Chui, Loffler and Roberts 2010). In a consumer setting, home gas, electric, and solar energy usage and production can be monitored and adjusted automatically for significant energy and cost savings, for example to avoid peak gas and electric rates. HVAC, lights, and refrigeration units can be set for lower power usage or turned off when no human presence is detected or expected. Electric vehicles can be charged when electric power is most cheaply available. The efficiency of automatic braking or collision avoidance systems in automobiles can be improved as sensors and feedback become more advanced.

The most demanding use of the IoT involves rapid, real-time sensing of unpredictable conditions and instantaneous responses guided by automated systems mimicking human reactions (Chui et al 2010). For comparison, one might consider the rate of data the human eye sensors record (perhaps megabits per second), transmission of an appropriately reduced amount of optical information to the human brain, and the complex function and processing utility of the human brain in this automated process. All aspects (sensing, data reduction, process, archival storage) of the complex process of the human process of seeing will need to be better understood in the new era of the future IoT. Advances in robotics and artificial intelligence will be as important as efficient interoperability needed for a self-organizing IoT.

With automation comes the necessity of minimizing manual interventions, which is a security issue. How can one monitor all of these automated processes if they are being performed automatically without human intervention? The amount of software that will be needed to accomplish these security goals is exponentially larger than presently, also implying an exponential increase in software vulnerabilities. Will all automated systems be tested fully before they are released for public use? Since the systems will likely be dynamically created by a plant operator or home user, the answer is probably negative. Systems for highly critical processes may be better tested for vulnerabilities, but the general mode of vendors has been to release when functional requirements of a product are met, and worry about security later. This implies that our original automated IoT will be severely insecure (cf. Green 2015).

## Robotics

Industrial robots have the potential to change production processes as much as computers have changed the office work environment. Robots can be designed for performing operations quickly, repeatedly, and accurately. They have applicability across many different domains in the manufacturing industry and have added tremendous value to various manufacturing processes. Petro chemical industry, for example, has used robotic systems to improve safety and efficiency, and to reduce environmental impact. In regions where it is difficult or dangerous for humans to work, robots may be enabled to carry out such tasks as maintenance, inspection and repairs (Heyer 2010). As robots are introduced to these types of environments, however, issues of trust and accountability come into consideration. One must also consider how the robots will fit into the organizational structure. Finally, any distributed system introduces vulnerabilities in the network layer. These vulnerabilities can be compromised in such a way as to sever or corrupt communications. They are also susceptible to all the D5 effects noted in the previous section.

Some robots are built to operate autonomously, with little to no human intervention, and some are remotely controlled. In order for the next generations of users and operators to trust autonomy, however, it must be predictable enough to operate under complex and dynamic conditions with high confidence levels and still be able to be tightly controlled or potentially instantly interrupted by the human operator (Murphy & Shields 2012). Maintaining this flexibility in future system will allow for sufficient levels of confidence in the actions performed by our robotic counterparts.

The human response to increased levels of autonomy also needs to be considered. If robots have too little autonomy, human operators will waste time attending to robots instead of attending to their work tasks. Also, a new skill set will need to evolve for future human operators if they are going to be skilled enough to fix or maintain robots in their manufacturing environments.

The main benefits of autonomous capabilities are to extend and complement human performance, not provide a direct replacement of humans. If robots are highly autonomous, situational awareness of plant activity may start to diminish (Kott, Wang, Erbacher; 2014). Robots can augment human perception, action, speed, persistence, resistance to fatigue. They can permit delegation and reduction of cognitive load. Some robots will be equipped with the ability to perform inspection and sample taking, while others will carry out more sophisticated operations like maintenance and repairs. Together, they can enable operation in areas too hazardous for humans to work in. (Heyer et al. 2010).

Some experts advocate that no matter how much we depend on robots and autonomy, we should ensure humans have ultimate control. Humans need to oversee, and have the ability to modify behavior as needed. As our trust in robots and autonomous systems increases, the range of levels of autonomy available can shift over time as needed (Endsley 2014)

In situations where the work space is dangerous for humans, robots can be used to improve safety in the workplace. Robots are not as vulnerable to workplace hazards including high temperatures, hazardous chemicals, radiation, and reaching difficult physical access points in manufacturing environments. Mobile robots including unmanned aerial vehicles have been developed to work in disaster response, inspections of infrastructure and decommissioning of nuclear plants. A key technology for the robots is teleoperation that enables humans to control robots remotely. (Hirukawa 2015)

Autonomously guided vehicles have been widely used for manufacturing, mainly for carrying parts in factories, and in other applications of robotics for logistics. Robots are also used in manufacturing facilities today to unload and move parts from trucks to the plant supply rooms while simultaneously maintaining inventory accountability and control. This role of robot systems is likely to increase in years to come.

There are other noteworthy types of robotic systems that are gaining popularity in manufacturing; robotic human augmentation and nano-bots. These are two areas are worth discussing because there are being extensively researched in the defense and security fields today.

### Nanobots

Nanobots are a type of microscopic robot. A nanorobot is any artificial machine with overall size on the order of a few micrometers or less in all spatial directions and constituted by nanoscopic components with individual dimensions in the interval between 1 and 100 nm (Requicha 2003).

A nanobot device has shown to have the capability to move quite freely through the entire human body circulatory system. One can envision a future where these nanobot technologies could be used in a manufacturing process, for example, to provide a microscopic view into the process conditions critical to certain bio-pharmaceutical or nuclear facilities.

The idea of surveying the state of fluid suspension with swarms of nanobots could be demonstrated in the bloodstream.  A nanobot in a capillary has demonstrated the ability to feel the metabolic pattern of the family of cells fed by the capillary itself, thus surveying the cells contained within a given length of the tube.  Each nanobot is a self-propelled machine, obtaining energy from the environment, and is able to recognize and dock to the components within their process (Cavalcanti et al. 2006). They can sense membranes and subsequently recognize the state of health of its environment.  They also may be used to store the information, to transfer it to the central unit, and eventually take actions which may have an effect on the overall process conditions.  Within a swarm of nanobots, each bot stores specific chemicals to be released for detection by other nanobots (Cavalcanti et al. 2006).  This could also be used in a manufacturing setting to transfer information from one location in the process to the other.

Ensuring that nanobots and nantobot swarms are operating securely is a complicated matter.  Nanobots are by definition extremely small and are therefore very difficult to monitor for individual malicious behaviors, especially if a large swarm of nanobots is deployed.  If individual nanobots are programmed with software, how might one scan the nanobot operating code for infections?  If nanobot swarms are programmed with chemical means, would there be a means to ensure that the function and control of the swarm not be overtaken by a malicious actor, in the same manner that viri and bacteria affect human biological receptors?  How will the health monitoring and maintenance of the nanobot swarm be performed?  When nanobots reach the end of life, how are they disposed?  As with other aspects of innovative IoT devices, nanobot systems offer incredible utility but have not been yet designed or analyzed for safety and security.

*Software*

Software and Applications

Getting all segments of the IoT to communicate and work together is key to its success.  This means deploying significant volumes of the software and middleware that will enable the diverse hardware devices to talk to other hardware and the IoT infrastructure (Karimi and Atkinson 2013).  Much of the software will be local to the devices and will be provided by the vendors of the hardware devices.  Because the devices are inexpensive and easily replaced or upgraded, software patching for security or other purposes will likely be neglected or ignored, especially by consumers.

IoT solutions do not follow a unified business model (cf. Schartel 2015), and over time software engineers and architects will need to accommodate the requirements of additional diverse stakeholders.  Currently, security guidance and technical guidelines for global interconnectivity are poor and incomplete.  There is not a clear understanding of preferred methods for how devices will identify and automatically interconnect to local networks and cloud data services, let alone how they will do this in a secure fashion.  Since the IoT market is driven by vendor markets, cooperation will be needed by major vendors to establish guidelines and requirements for software engineers who write code for vendor devices.  Consumers and industry owners will need to demand increased authentication security and reliability, especially for IoT components of critical control systems.

The software that makes IoT devices "smart" will have varying levels of "smartness." Efforts to add "smartness" to devices will be popular for some things with IoT connectivity, specifically, things with longer life cycles. Examples, include local networks in automobiles, large home appliances such as refrigerators and televisions, home lighting and home security systems, and most industrial control system components that were never designed for the IoT. As the IoT matures, software engineers will be able to accommodate requirements for security and interconnectivity between multiple vendors' things. Currently, however, most vendor business models seem to focus on producing products quickly for maximum profit, and to neglect security features until they are demanded.

The future IoT will also generate tremendous amounts of new sensor data and information. Markets for software for data management, data formatting, data storage, and secure data transfer will boom as the size of the IoT grows. Methods for ensuring data privacy will be demanded by the consumer, but methods for data mining will also generate an increased software demand as corporations realize the potential for profit optimization from the new IoT information. Network infrastructure usage patterns and personal information not protected by privacy laws will be harvested and offered for sale by those providers with the most intelligent software products. A new layer of compliance software may be needed to ensure that government privacy laws are enforced. Software analysts working for financial firms will turn their attention toward the new IoT data and will develop tools for market prediction.

System automation, artificial intelligence, and automatic network and device authentication are integral to the IoT, yet they are non-trivial problems solutions to which are not yet fully developed. Ensuring that automation occurs is vital to the development of the IoT. Ensuring that automation is secure is vital to sustainment of the IoT. The level of software effort needed for automation is tremendous, not only because the number of devices is increasing exponentially, but because there will be a continuous need for requirement definition and redefinition as the IoT architecture begins to be affected by all of the stakeholders. Software will need to be continually revised, rewritten, and reused to accommodate the changing requirements. Lack of attention to changes in the software will create software vulnerabilities in device and network access, cloud and data storage, and any other IoT component.

Software apps will take a different approach in the IoT context. In the current approach, users use a few apps every day for everyday tasks. IoT device manufacturers will not be able to provide a single app for controlling their unique function, since users will not be able to accommodate a huge number of these simple apps. There will need to be a consolidated effort to provide the consumer (e.g., cell phone user) or industrial control system operator with apps or software that monitor and control a large number of device functions. Such an app will need to condense the information and provide some level of security alerting when device values need attention. This software will need to be universal in the sense that it can accommodate a new type of IoT device to which the consumer or operator would like to connect. The software needs to accommodate the device in an automated fashion, since from a practical standpoint, the user will not be able to download vendor software each time a new device connects. Semantic middleware for the IoT (cf. Whitmore, Agarwal and Da Xu 2014) may offer a solution for this problem.

As with most of the IoT, functionality of these software systems will be the initial focus, and security will be a secondary consideration. What is important, however, is to realize that as any system (such as the IoT) grows in complexity and intelligence, the dependence on software increases, and software, being a human product, has imperfections. Incorporation of greater automation into the system also means there is less inspection by humans. Computer-aided tools will need to allow reliable security monitoring of this complex system. If the future IoT is to be safe for the consumer and industry, improved security methods will be needed. The attack surface presented to an adversary will be exceedingly large if one scales current interconnected devices to IoT scales and makes them all Internet accessible. Network isolation and segmentation with virtualization and hardware-based security methods (e.g. Ukil, Sen and Koilakonda 2011) may help.

## Users of the Future IoT

We discuss two distinct groups of users of the future IoT: users of future industrial plant control systems (i.e., the IIoT), and consumers who will use the larger scale IoT.

### Industrial Plant Users

Cyber-attacks in manufacturing environments are becoming more sophisticated, leveraging remote access vulnerabilities, supply chain interdiction, and insider threats. In the next three sub-sections, we discuss key aspects of the IIoT that will be affected by its ongoing evolution.

### Plant Control Methods

The first control systems were mechanical and integrated in one mechanism the sensor, the actuator and the controller. For example, in the speed regulator invented by James Watt the centrifugal force exerted on two spinning masses moved the lever that controlled the flow of steam to the engine. That enabled a proportional-only control.

Pneumatic and hydraulic control systems were first developed for ship steering in the 1890's and soon after were applied to manufacturing (Bennett 1996). Through various types of physical devices operated by compressed air or hydraulic fluid, it was soon possible to perform proportional, integral and derivative control (PID). Until the introduction of electronic controls in the second half of the 20$^{th}$ Century, most manufacturing automation used stand-alone single loop pneumatic and hydraulic controllers. Multi-variable control required complex assemblies of physical devices and tubing and a change in control strategy necessitated changes in the tubing and often new devices. Tuning was done in the field controller with knobs.

Much simpler solid-state analog electronic sensors and controllers were introduced in 1959 and spread rapidly, while the motive force for actuators generally remained pneumatic or hydraulic. At that point, changes in control strategy required only rewiring and installing inexpensive components. The first digital and freely programmable control systems were introduced in 1969, replacing the traditional hardwiring of analog logic and control programs (Kruger et al. 2014). However, the functionality remained mostly the same PID control as in the original pneumatic and hydraulic devices due to the tendency towards one-to-one replacement and the availability of well-established methods for PID loop tuning and troubleshooting (Bennett 1996).

There is no particular reason to use only PID control, as other control strategies can be programmed, such as RST (Discrete-time linear MISO controllers), SFO (State Feedback and Observers), MPC (Model Predictive Control) and Fuzzy Logic Control. Of those, the most successful has been MPC, typically used in supervisory mode with PID controllers at the base level. MPC offers drastic improvements in set point responses for multivariable systems because of the coordination it provides (Astrom & Hagglund 2001).

Yet, the bulk of the industrial control systems are single PID loops. Often they are not performing as well as they could. In a typical plant 50 out of 100 PID loops will show degraded performance after six months. Typically, 30% of the loops are run in manual mode, 15% have an output out of range, 30% are increasing process variability instead of reducing it, and only 25% are actually improving the process (Starr 2015). The increased automation and optimization promised in the IIoT will help improve the efficiency of these control loop processes. Since the automation will be under software control, it will be necessary to analyze and monitor access and use of that code in order to maintain secure and safe operability of the added automation.

### Data Transfer Media in Plants

Data transfer media is also evolving. Older process plants were built with 2-wire twisted-pair cable networks, connecting all the process units and measuring instruments together in an overall plant control scheme. These relatively unsophisticated instruments convert their measurement by various means into a 4-20 mA output or pulse signal to the control system. The more advanced technologies, such as Coriolis, ultrasonic or electromagnetic flowmeters have, until 2006, required a dedicated power supply for their functions, in addition to the output loop, and thus a 4-wire infrastructure was required as a minimum. Newer flowmeters can also be installed with a single 2-wire connection, and the low energy levels supported by these 2-wire loops are more easily rendered safe, in terms of explosive risk in hazardous areas containing flammable materials. However, the amount of information that can be passed back and forth is very limited.

### Smart Sensors

Many new sensors are revolutionizing the manufacturing process already. A smart sensor often contains a microprocessor that conditions the signals before transmission to the control network. It filters out unwanted noise and compensates for errors before sending the data. Some sensors can be custom programmed to produce alerts on their own when critical limits are reached. Caution needs to be taken, however, to ensure such sensors have the proper security protocols in place to prevent a cyber intruder from tampering with the controls.

In contrast, the soft sensor, or virtual sensor, is a piece of software which represents a "sensor" that is not actually there. Often this sensor "output" considers several physical sensor values and fuses the data together to provide a new sensor value. The soft sensor may represent dozens or even hundreds of measurements. Soft sensors are especially useful in situations when the insertion of a physical sensor is not feasible. Software algorithms that are used to generate the output values of soft sensors include Kalman filters and Artificial Neural Networks. As with other new software-based features in the IIoT, one must be aware that additional software control introduces additional avenues for malicious manipulation of system control processes, and appropriate security measures need to be taken to ensure that unauthorized remote access and other vulnerabilities are mitigated.

## The Network Layer

EtherNet/IP Standard IEC 61784-2 is open, manufacturer-independent and stable, and is supported by more than 300 member companies and hundreds of products. Using EtherNet/IP allows the user to access to the smartness of multivariable devices (Endress+Hauser 2014). For example, data regarding mass flow, density, temperature, totalizer settings as well as diagnostics can be delivered over a single cable. In addition, savings of 40% can be made through reduced commissioning time. The time spent on loop identification, device integration and process-loop tuning can also be reduced by 25%.

The connected factory provides a clear set of architectural guidelines and products that tie together factory automation systems, enterprise applications, and the wider ecosystem of supplier and partner solution. The common architecture will be more scalable for ruggedized Industrial Ethernet and enterprise networks. It will offer a standards-based Industrial IP Ethernet switching and security service.

However, as things become more connected, the cyber-attack surface and the vulnerabilities opened up through the increasing number of access points becomes a greater concern. Monitoring systems which employ behavior based analysis in industrial control systems are gaining more popularity as it is becoming more difficult to rely on threshold based or single- point of failure based alerting. (Blowers 2014). There will be a need in future systems to have an autonomous supervisory system to monitor the overall process behaviors of the manufacturing processes.

As shown in past cyber-attacks on industrial systems, like STUXNET, it is becoming quite common to target single control loops or spoof specific sensor outputs. Monitoring systems which employ behavior based analysis of events occurring in the industrial process will need to be integrated with the network layer so that correlations and dependencies may be baselined, and anomalies can be quickly detected (Blowers 2014).

### Consumers

Arguably, the most important long-term influence on the construction, security, and support for the future IoT is the consumer. Wearable devices and home controllers are built by vendors at the demand of consumers. New innovative connectivity devices and entertainment features for automobiles and home are created due to consumer interest and demand. Wireless connectivity and other convenience technology eventually migrates into industrial settings as plant owners realize lower costs and easier (but often less secure) operations. Consumer cell phones are no longer only portable communication devices – they are now centralized access points for much of information. The philosophy of interconnected apps used on a cellphone is a primary inspiration and basis for the IoT.

The security issues associated with consumer demand for innovative IoT are enormous. They are also unknown. Vendors will continue to develop and market new devices and sensors that they believe consumers will want. They will not generally be designed for security, although current concern about data privacy may be addressed at some level. The ability to create secure IoT devices and services depends upon the definition and agreement of security standards for the anticipated interconnectivity methods. Until the methods are defined and the security issues are addressed by all vendors, new devices and sensors for consumers will seriously increase the vulnerability of the IoT as it develops (e.g. Green 2015).

Unlike IoT, the future IIoT will benefit from the fact that adoption of new devices and methods will come slower since control engineers will be resistant to potentially dangerous new technologies. Some security issues that will be adopted by consumers in the larger scale IoT may be resolved by the time they are adopted in industrial control settings. However, it may be impossible for control engineers to have enough time to adequately evaluate these new technologies before they have to be adopted.

## Support for the Future IoT

In this subsection, we discuss services and collaborative efforts for supporting users of the IoT.

### Computing and Infrastructure

#### Industrial Control Efficiency

Efficiencies are already being realized in manufacturing and even in our own homes with the inclusion of industrial control systems for everything from heating, ventilation, and air conditioning control to ambient light sensing and adjustment. These capabilities allow facility operating costs to be slashed by adjusting temperature and lighting based on occupancy. In addition, through data collection, trends for energy consumption can be developed and monitored to support problem diagnosis. However, this is just the beginning of the Internet of Things revolution in manufacturing. From a facility maintenance perspective, smart devices such as emergency lighting and smoke detectors can alert maintenance staff proactively when problems occur. Mundane tasks like monitoring soap levels in washrooms can also be automated to reduce staff levels and decrease response time. Other technologies such as smart elevators promise to more efficiently manage resource use and minimize wait times for users by predicting peak usage and positioning cars strategically for response.

A larger increase in efficiency will come when smart IoT products begin communicating between themselves automatically. For example, infrared and motion sensors could communicate to other systems that there has been no human activity in the home or office, and thus appliances such as water coolers, HVAC systems, and water heaters could be switched to a lower-power standby mode. This obviously happens with no human intervention and presents a security issue. Automatic authorization and command of utilities and appliances can be dangerous if it is possible that they can be set to unsafe ranges. The communication method between smart devices must be secure so that outsiders will not have access to information such as when humans are present or not.

#### Networks and Infrastructure

The future IoT will require significant changes in supporting infrastructure to accommodate the increased number of addressable sensors and devices, and the diversity in how those devices communicate. It is however unclear what architectural changes will occur, since relevant interoperability guidelines, communication standards, and vendor designs are still immature.

IoT end devices and sensors currently use a large number of communication methods, such as Bluetooth, NFC, RFID, ZibBee, WiFi, Ethernet, and cellular protocols.  The TCP/IP 3-way handshake produces unwanted overhead network traffic for some inter-device packet communications, and may need to be replaced for some communication links in the future IoT.  Requirements for global addressability of things for IoT automation may be satisfied by using IPv6 addressing.  An alternative may be to abandon global addressability using NAT or local-only addressing such as private IPv4 addressing.  This may provide an additional layer of security over a globally-accessible IoT model using IPv6.  IP addressing is not ubiquitous among IoT devices – many use RFID, Bluetooth, or proprietary addressing methods.  Some researchers in the field of future IoT network architecture propose integrating technology from IEEE 802.15.4 wireless sensor networks (WSN) with that of RFID systems and IP networks (e.g. Castelanni et al. 2010, Atzori et al. 2010, Gubbi et al. 2013).   This may allow use of small packet frames compared to what is needed for the IP protocol.  Most wireless sensors using WSN spend most of their time "sleeping" so that they are not responsive.

Clearly the initial IoT will be a blend of many different types of hardware using diverse communication methods.   This will increase the attack surface for malicious outsiders, and is likely to create a privacy issue for new systems designed without security in mind.

In time, the future IoT networks will be self-aware and self-adapting, so that they can accommodate the bandwith and global connectivity requirements of new subnets of things when they connect.  This will require better standardization and interoperability guidelines between vendors so that the networks can accomplish efficient transport and maintain network security of the data on the subnet devices.  Research in network management methods and secure software-defined networking is needed.

A careful accounting of the IoT devices should be accommodated by the infrastructure, similar to mobile technology.  As proposed by Zorzi et al. (2010), the infrastructure should:

- Discover entities based on identifier, location, type, provider, etc.

- Provide a lookup service for entity properties, which would allow interaction with the device

- Monitor the state of the entities, and keep the lookup information and links up to date

It is obvious that the desired state of the future IoT infrastructure is quite different from the present state, and that much research and development is needed.  Security must be considered and tested as part of that effort.

### New Territories for Network Complexity

As the IoT network infrastructure grows, as expected, in the exponential manner, an even more fundamental environmental game changer may occur. We will eventually cross a network complexity threshold and enter new territories beyond the limits of conventional system manageability, perhaps even stretching human comprehension. Qualitative increases in technological complexity—enormous in size, connectivity, interdependence, heterogeneity, and dynamic capabilities—coupled with the exploding network growth occurring now in under-served communities worldwide might defeat conventional scientific and engineering approaches to cybersecurity.

Right now, the cyber-research community offers few insights to help us observe, stabilize, and control very-large-scale and multidimensional networks. There is still much for us to understand about how social-cognitive and cyber-physical links will govern overall network complexity. Even single vendors have problems keeping up with all of the items in their product line. We expect vendors to produce large numbers of inexpensive devices with short lifetimes on the order of less than a year. Those devices will be present on the IoT long after they are no longer supported by the vendor. Nobody will fully understand the devices in their network. This increased system complexity enhances opportunities for adversarial attack.

## Computing and Cloud Services

Computing and data storage methods have changed drastically from centrally located large single mainframes in the middle of the century to powerful desktop computers and servers in the later part of the century. Further, a client-server service model has emerged so that much of the computing and storage is done remotely and a thin client interacts with the remote service over the Internet, or "the cloud." The current cloud model will necessarily support initial IoT devices since this is the model in place, but some changes will be needed to accommodate the expected architecture of the IoT.

As mentioned, devices in the IoT may not be globally addressable via an IP address and may not be directly connected to cloud storage and processing. In addition, one need not keep all of the data coming from a sensor device. The data needs to be processed locally into useful, intelligent information, which can then be forwarded to a proxy device such as a cellphone or dedicated server. The proxy device can then store the data locally or push it to the cloud for further processing. Privacy concerns can be a factor in protecting the sensor data. There is no doubt that cloud services will need to scale with the growth of the IoT, but it should not scale exactly to the size of the data collected by the sensors. Some of the sensor data will be thrown away, and some will be sent to a central cloud for storage and further processing. Restricted-access local cloud storage and processing may be useful for temporary or permanent sensor data, especially when sensor data or cloud service traffic becomes prohibitively high to push to a globally-accessible public cloud. Gubbi et al. (2013) propose a scalable cloud framework which allows networking, computation, storage and visualization themes to scale separately, accommodating the indeterminate growth of the future IoT.

Context-aware computing (e.g. Perera et al. 2014) will be important as IoT sensor data volume becomes large and data owners wish to better harvest the value of the information. The data collected by the sensors will not have high value unless it is properly understood by storing context-related information with the raw sensor data so that data interpretation is more meaningful. Cloud storage and processing techniques can then be used to analyze additional contextual meta-data together with the sensor data. Examples of context-oriented meta-data are location and time information, data owner, digital chain of custody, access information, and medical history (for health care data).

An important aspect of future IoT cloud storage and computing is that ISPs and telecommunication companies control access to the data, and may even have preferential rights to data the customers store on their platforms. Once your data leaves your globally-connected IoT sensor or IP-connected proxy server for the cloud, you no longer have the ability to secure the data. You must encrypt the data or rely on provider security. While the providers may mean well and may have very high security standards, they cannot provide 100% protection against unauthorized access.

### New computing paradigms

While standard silicon-wafer CPU computing methods are commonplace today, emerging computing paradigms—nanocomputing, quantum computing, biologically or genome-based computing—might develop soon enough to make most current cybersecurity technologies obsolete, thus drastically changing the related markets.

Quantum computing and networking are already fueling lively debate, with one side making claims for the technologies' inherent security while the other side highlights the opportunities it presents for hacking. Biologically inspired computation and communication paradigms—for example, the Gaian dynamic distributed federated database (Toth et al. 2013) and related cybersecurity applications, such as artificial immune systems—will attract growing interest, especially as they offer promises for autonomous adaptation to previously unknown threats and even self-healing (Kott 2014).

If implemented, these emerging computing methods would bring an exponential layer of complexity to the IoT. Security and privacy of data would be unpredictable for systems that rely on strong encryption. Depending on the cost of the computing methods, centralized processing with cloud systems could become obsolete, especially if IoT sensor processing power both increased dramatically in capability and remained low cost.

### Government and Industry Guidance and Collaboration

The exact functional nature of the future IoT for both industry and consumer cannot currently be defined. Regulators and interoperability collaborations will drive the development of future cross-industry standards (McDonald 2014). While the development of IoT standards is underway, larger organizations may not have an interest in participating if their IoT market share does not seem threatened. Gartner estimates five billion smart devices will be in use by the end of this year (2015), and yet no central IoT standards are in use, and there is no real oversight of IoT development methods (Null 2015). It is clear that a lot of work will be needed to develop a large number of standards to make the IoT function efficiently and safely, but as asserted by Schneier (Green 2015) at the moment "it's all really, really bad and it's going to come crashing down."

Some examples of current IoT standards groups are Thread, AllSeen Alliance, Open Interconnect Consortium, Industrial Internet Consortium, the ITU SG20 standards group, the IEEE P2413 project, the Apple HomeKit, the IETF RPL, CoAP, and 6LoWPAN protocol standards groups (Null 2015, Sheng et al. 2013). These cover a wide range of technical issues, such as M2M communication, interoperability between large vendors, wireless communication standards, home and user based technical IoT issues, addressability and routing issues. It is a good start, but it is only a start.

Better solutions will need to be developed as the IoT builds out and more vendor devices and functions need to be accommodated. There seems to be a lack of standardization effort related to data models, ontologies, and data formats to be used in IoT applications; this may present a barrier to innovative development of key IoT technologies (Miorandi et al. 2012). The rapid growth of the IoT makes efficient standardization difficult if not impossible. Specific issues in IoT standardization include vendor interoperability, radio/wireless access, security and privacy, addressing and networking, and guidelines for industrial environments (Da Xu et al. 2014, Atzori et al. 2010). Efficient allocation of the wireless spectrum by the FCC and similar organizations will be needed if the future IoT is to have the envisioned wireless interconnectivity. Interoperability agreements and standards and vendor collaboration will take some time as IoT market leaders engage with each other and IoT users. Since IoT development is market-driven, there is no single architect to organize this effort. Governments can provide some guidance, but cannot regulate the future IoT any more than they can regulate the global Internet.

The importance of governance in ICSs is discussed in an earlier chapter of this volume which mentions that while unstructured short—term successes are vital, long-term success requires a more structured approach. Stakeholders are less interested in making informed decisions toward an overarching plan when environments are increasingly connected (Westby 2003). Governing for security means viewing adequate security as a nonnegotiable requirement of being in business (Allen 2005). The governing body must have the authority, accountability and resources to act and enforce compliance. Among the governing documents within an organization, the most powerful to enable resource decisions and revealing to make security modifications are assessments. Four examples of current assessment methods for control systems security are NIST Cybersecurity Framework, DoE C2M2, RIPE Framework, and the DHS CSET framework. Each approach is based on years of subject matter experience and community best practices. The amount of experience securing the IoT (and future IoT) is obviously significantly less than that for current control systems, which explains in part the current inability to propose useful governance for the IoT. Not only is there no functional architect, there is also no security architect, or governing body.

## Predictions and Potential Solutions

The future manufacturing will evolve to accommodate many global changes. There will be limitations on resources such as energy, population, special metals, etc. The population will demand more products. In the US there is a strong decline in the number of students pursuing education in science, technology and math, and a decline in the number who are willing to pursue career fields in manufacturing. This may significantly increase the demands for autonomous systems and robotics.

The interconnectivity of things is creating a world of unknown potential,. Through distributed systems, information sharing is greatly improved. Information can reach a wider population, and products and services can be made more readily available. However, it also makes us significantly more vulnerable than we ever could have imagined in the recent world of isolated systems.

In the following subsections, we envision some potential solutions to the anticipated security challenges in the future IoT (Kott, Swami, McDaniel; 2014).

### Resilient self-adaption

Potential innovations based on resilient self-adaptation could be very important for the security of the future IoT. Cybersecurity in this case will derive largely from system agility, moving-target defenses, cybermaneuvering, and other autonomous or semi-autonomous behaviors (Jajodia et al. 2011). Exploiting such self-adaptation might mean shifting a significant fraction of design resources from reducing vulnerabilities to increasing resiliency.

A truly resilient system could experience a major capability loss due to cyberattack, but recover sufficiently rapidly and fully so that its overall mission proceeds successfully. For example, promising results have been shown for software residing on a mobile phone to perform self-healing—by applying patches or self-rewriting code—in response to abnormal behaviors it detects (Azim et al. 2014).

However, effective autonomous self-adaptation calls for a degree of machine intelligence far ahead of what's now imaginable and would also increase system complexity, thus multiplying vulnerability risks. Given that complex attacks, along with their circumstances, are both diverse and unpredictable, achieving practical resiliency is no more than probabilistic—not a comforting thought for future systems operators.

### Mixed-trust systems

New design methods for mixed-trusted systems may also be important for future IoT security. We see these as security-minded, flexible, modifiable systems that combine and accommodate untrusted hardware and software—resulting from dubious supply chains, legacy elements, accreted complexity, and numerous other sources—with clean-slate components. Related ideas include a management protocol that applies trust-based intrusion detection to assess degrees of sensor-node trustworthiness and maliciousness (Bao et al. 2011).

Success depends on qualitatively significant changes in the design methodologies and tools that enable complex systems to be synthesized--for example, reinforcing untrusted components with clean-slate, highly trusted "braces." Such designs would also have to include components that could be rapidly and inexpensively modified to defend against new threats as they are discovered. A breakthrough in current formal methods or the emergence of as yet unknown but highly reliable semiformal methods would thus be required.

### Big Data Analytics

Though still immature from a cybersecurity perspective, big data analytics—predictive and autonomous—is an area already exerting a noticeable influence. Potentially reaching global scale, able to anticipate multiple new cyberthreats within actionable timeframes, and requiring little or no human cyberanalysis (Gil 2014), big data analytics is a game changer that could bring new potency to cyberdefense.

Much of this power will likely derive from aggregating and correlating a broad range of highly heterogeneous data, which is challenging in itself. Add to this heterogeneity the noise, incompleteness, and massive scale characteristic of cyberdata, and the challenges only increase (Kott & Arnold 2013). Much work remains for developing algorithms that can ferret out deeply hidden, possibly detection-protected information from so heterogeneous a mass.

### Proactive Threat Responsiveness

Finally, IoT security may be improved through the possible emergence of proactive threat-source responses: strategy-oriented approaches, offense-based techniques, alternative security postures, and deception- and psychology-aware mechanisms. Currently, little is understood about the shape such methods might take, especially in view of the legal and policy uncertainties surrounding cybersecurity in general, and proactive cyberthreat responses in particular.

Extensive strategic and tactical knowledge developed through our long experience with conventional conflicts might offer important insights about anticipating adversaries' actions (Ownby and Kott 2006), holding adversaries at bay and defeating their will to attack. But focus on the past might also mislead and limit our thinking.

Whatever the details, any such approaches will benefit from greater situational awareness and require understanding our adversaries' architectures, infrastructure, and sensing capabilities, as well as we do our own. We will also need languages to help clearly and precisely articulate the specific defensive and offensive circumstances, cultural intelligence and adversary modeling, and deep insights into individual and collective cognitive processes.

## Summary and Conclusions

IT and control systems manufacturers are seizing the opportunity of having new novel hardware devices as the "Internet of Things" begins to scale up. As the number of devices continues to increase, more automation will be required for both the consumer (e.g. home and car) and industrial environments. As automation increases in IoT control systems, software and hardware vulnerabilities will also increase.

In the near term, data from IoT hardware sensors and devices will be handled by proxy network servers (such as a cellphone) since current end devices and wearables have little or no built-in security. The security of that proxy device will be critical if sensor information needs to be safeguarded. The number of sensors per proxy will eventually become large enough so that it will be inconvenient for users to manage using one separate app per sensor. This implies single appls with control many "things," creating a data management (and vendor collaboration) problem that may be difficult to resolve. An exponentially larger volume of software will be needed to support the future IoT. The average number of software bugs per line of code has not changed, which means there will also be an exponentially larger volume of exploitable bugs for adversaries.

Until there are better standards for privacy protection of personal information and better security guidelines on communication methods and data/cloud storage, security of wearable and other mobility devices will remain poor. More work needs to be spent on designing IoT devices before too many devices are built with default (little or no) security.

Physical security will change as well. As self-healing materials and 3D printers gain use in industry, supply-chain attacks could introduce malicious effects, especially if new materials and parts are not inspected or tested before use.

The main benefits of autonomous capabilities in the future IoT is to extend and complement human performance. Robotic manufacturing and medical nanobots may be useful; however, devices (including robots) run software created by human. The danger of the increased vulnerabilities is not being addressed by security workers at the same rate that vendors are devoting time to innovation. Consider how one might perform security monitoring of thousands of medical nanobots in a human body.

The ability to create secure IoT devices and services depends upon the definition of security standards and agreements between vendors. ISPs and telecommunication companies will control access to sensor data "in the cloud" and they cannot provide 100% protection against unauthorized access. IoT user data will be at risk.

Diversity of the hardware and software in the future IoT provides strong market competition, but this diversity is also a security issue in that there is no single security architect overseeing the entire "system" of the IoT. The "mission" of the entire IoT "system" was not pre-defined; it is dynamically defined by the demand of the consumer and the response of vendors. Little or no governance exists and current standards are weak. Cooperation and collaboration between vendors is essential for a secure future IoT, and there is no guarantee of success.